# X-ray Variability of Cygnus X-1 In Its Soft State

W. Cui

Center for Space Research, Massachusetts Institute of Technology, Cambridge, MA 02139

S. N. Zhang

NASA/Marshall Space Flight Center, Huntsville, AL 35812

K. Jahoda, W. Focke, J. Swank

NASA/Goddard Space Flight Center, Greenbelt, MD 20771

W. A. Heindl, R. E. Rothschild

Center for Astrophysics and Space Sciences, University of California, San Diego, La Jolla, CA 92093

## ABSTRACT

We reported previously that for Cyg X-1 there is a settling period following the transition from hard to soft state (Cui et al. 1997). During the transition, The low energy spectrum (below $\sim 10$ keV) varies significantly from observation to observation while the high energy portion changes little. The source reaches nominal soft-state brightness during the settling period. It can be characterized by a soft low-energy spectrum and significant low-frequency 1/f noise and white noise on the power density spectrum (PDS). The low-energy spectrum becomes even softer, and the PDS is completely dominated by the 1/f noise, when the "true" soft state is reached. In this paper, subsequent *RXTE* observations of Cyg X-1 in the soft state are examined, and the results confirm our earlier conclusions. We also present the results from the observations taken during the soft-to-hard transition. As expected, the white noise appears again, and accordingly the 1/f noise becomes less dominant, similar to the settling period at the end of the hard-to-soft transition. The low-frequency 1/f noise has not been observed when Cyg X-1 is in the hard state, so it appears to be positively correlated with the disk mass accretion rate which is low in the hard state and high in the soft state. The difference in the observed spectral and timing properties between the hard and soft states is qualitatively consistent with a simple "fluctuating corona" model (Cui et al. 1997). Here we present more evidence for it.

## 1. INTRODUCTION

Despite decades of extensive studies, discrete spectral states in black hole candidates (BHCs) continue to puzzle us. These states, as well as the transitions between them, have been observed in several systems on many different occasions (see a review by Tanaka & Lewin 1995). For a given source, the states seem to be repeatable, and each displays distinct spectral and timing properties. A great deal of effort has been devoted to phenomenologically characterize them for known sources. As a result, our understanding of spectral transitions in BHCs has been greatly advanced, although we still do not know their physical origin(s). Similarities in different systems seem to hint a common origin. Therefore, deeper insights into one system will help understand the physical processes involved in others.

Cyg X-1 is the first binary system in our Galaxy that is dynamically determined to contain a stellar-mass black hole (Webster & Murdin 1972; Bolton 1972). It appears to have two spectral states. It spends most of the time in the hard (or low) state where the soft X-ray luminosity (2-10 keV) is low and the energy spectrum is characterized by a single power-law with a photon index of $\sim 1.5$ (Oda 1977; Liang & Nolan 1984). The X-ray flux varies on all time scales down to a few milliseconds, and the PDS can be characterized by a flat (or white) component with a low-frequency cutoff in the range of $\sim$0.04-0.4 Hz (van der Klis 1995 for a review). An additional steepening of the PDS above 10 Hz was also detected (e.g., Belloni & Hasinger 1990). Cyg X-1 has only occasionally been observed in the soft (or high) state (see reviews by Oda 1977 and Liang & Nolan 1984), and is not well studied in such state. The recent results from *RXTE* observations of Cyg X-1 in the soft state confirmed that its X-ray spectrum can be characterized by a hard power-law in both states, but in the soft state the low-energy spectrum ($< 10$ keV) is softer in the soft state (Cui et al. 1997), as observed previously (Oda 1977; Liang & Nolan 1984). The derived PDS for the soft state is dominated by the 1/f noise which breaks at some characteristic frequency and becomes $1/f^2$ (Cui et al. 1997).

Here we report preliminary results from subsequent *RXTE* observations of Cyg X-1 in its soft state and during a soft-to-hard transition. We have adopted the same data reduction and analysis procedure as described by Cui et al. (1997).

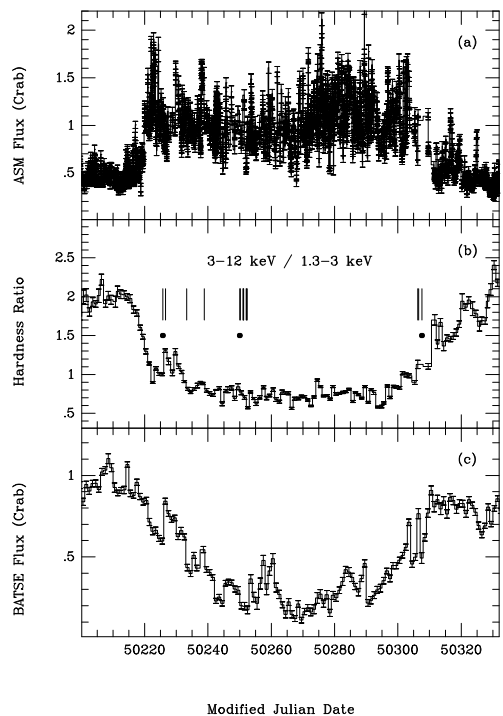

Figure 1: (a) The ASM light curve of Cyg X-1. It comprises measurements from individual "dwells" with 90-second exposure time. The vertical lines indicate when the RXTE observations were made, and the bullets show the ones that are concentrated on in this paper. (b) The daily-averaged time series of the ASM hardness ratio; and (c) The daily-averaged BATSE (20-100 keV) light curve. MJD 50220.0 is equivalent to 1996 May 17 0 h UT.

Shortly after the launch of RXTE, Cyg X-1 started a transition from the hard state to the soft state (Cui 1996; Zhang et al. 1996a). It spent nearly 2 months in the soft state, and started to go back to the normal hard state on 1996 July 26 (MJD 50290; Zhang et al. 1996b). The ASM (Levine et al. 1996) long-term light curve of Cyg X-1 (Fig. 1a) summarizes the entire period of the hard-to-soft transition, the soft state, and the soft-to-hard transition. Clearly, the spectral states should be taken seriously only in an average sense. Even in the soft state, the soft X-ray flux can occasionally drop to the hard-state level (see Fig. 1a). The source exhibits large X-ray variability in both states on time scales of hours to days. During the transitions, the ASM hardness ratio, defined as the ratio of the 3-12 keV count rate to that in the 1.3-3 keV band, shows a steady trend of spectral evolution (see Fig. 1b). When combining the results from the BATSE simultaneous monitoring of the source (Fig. 1c), a clear anti-correlation between the soft (ASM) and hard (BATSE) X-ray fluxes is firmly established (comparing Fig. 1a and Fig. 1c).

During this period, a total of 15 public Target-of-Opportunity observations (as indicated in Fig. 1b) were carried out with RXTE to monitor the temporal and spectral variability of Cyg X-1 in the soft state. The results from the first 4 observations were presented by Cui et al. (1997). Those from subsequent 8 observations taken during the soft state were also briefly reported to support our earlier conclusions (Cui, Focke, & Swank 1996). They will be presented and discussed in more depth here, along with those from the observations taken during the soft-to-hard transition. Limited by space, we have chosen to concentrate mostly on 1 (out of 8) observation in the soft state and 1 (out of 3) during the soft-to-hard transition (as indicated by bullets in Fig. 1b). For comparison, we have also included the very first RXTE observation to show the overall evolution of the spectral and timing properties of Cyg X-1 during both transitions and in the soft state itself. Table 1 summarizes the observation time and durations for these 3 observations. Only the results from the PCA observations are presented here.

Table 1: RXTE Observations of Cyg X-1

| No. | Observation Time (UT) | PCA Live Time (s) |
|---|---|---|
| 1 | 5/22/96 17:44:00-19:48:00 | 4208 |
| 2 | 6/16/96 00:00:00-00:40:00 | 816 |
| 3 | 8/12/96 14:40:00-15:58:00 | 2114 |

The RXTE mission is optimized for observing fast X-ray variability in a broad energy range. For the first time, true microsecond timing resolution is achieved (Bradt, Rothschild, & Swank 1993). The PCA has a $1°$ field-of-view (FWHM), and a total collecting area of about 6500 $cm^2$. It covers an energy range roughly from 2 to 60 keV with moderate energy resolution ($\sim 18\%$ at 6 keV).

## 2. SPECTRAL ANALYSIS

The observed joint PCA and HEXTE spectra were fit well with a model consisting of a blackbody and a broken power-law with high-energy cutoff (Cui et al. 1997). Because RXTE is still in the early stage of its mission, and the detector calibrations are still preliminary, we chose to take a cautious approach to spectral analysis by adopting this model, instead of more complicated ones proposed for Cyg X-1.

The model fits the observed PCA spectra pretty well, except for observations when the source reached the peak flux (at $\sim 1.5$ Crab), extensive discussions of which is beyond the scope of this paper. Table 2 shows the best-fit model parameters for the observations that are listed in Table 1. The uncertainties shown represents 90% confidence intervals. As reported by Cui et al. (1997), the best-fit $N_H$ values are all much larger higher than the interstellar value ($\sim 6.2 \times 10^{21} cm^{-2}$) (Bałucińska & Hasinger 1991). The stellar wind from the massive supergiant companion was proposed to be a possible mechanism that provides such large internal absorption. However, as pointed out, this result could be quite uncertain due to systematic uncertainties at the lowest energies. After all, the PCA cuts off at 2 keV, so we only see a small tail of the blackbody component. The observed spectra were fit again with the $N_H$ fixed at the interstellar value. The results are also summarized in Table 2 for comparison. Clearly, these fits have significantly higher $\chi^2$. To show what a typical soft-state PCA spectrum looks like, the observed spectrum for the second observation is plotted in Fig. 2a. The solid-line histogram represents the best-fit model. Fig. 2b shows the unfolded spectrum for the same observation with the contribution from each model component separated.

Table 2: Results of Spectral Analysis[1]

| No. | $N_H$ | $kT_b$ (keV) | $\alpha_1$ | $E_b$ (keV) | $\alpha_2$ | $\chi^2_\nu/dof$ |
|---|---|---|---|---|---|---|
| 1 | $13.8^{+1.5}_{-1.6}$ | $0.26^{+0.01}_{-0.02}$ | $3.04^{+0.05}_{-0.04}$ | $10.5^{+0.3}_{-0.3}$ | $2.03^{+0.02}_{-0.03}$ | 0.7/75 |
|   | 0.62 | — | $2.60^{+0.01}_{-0.03}$ | $10.3^{+0.2}_{-0.3}$ | $1.96^{+0.02}_{-0.02}$ | 1.4/76 |
| 2 | $13.3^{+1.3}_{-1.4}$ | $0.25^{+0.01}_{-0.02}$ | $3.22^{+0.05}_{-0.04}$ | $11.0^{+0.2}_{-0.3}$ | $2.06^{+0.03}_{-0.02}$ | 1.0/75 |
|   | 0.62 | $0.30^{+0.07}_{-0.08}$ | $2.80^{+0.03}_{-0.02}$ | $10.9^{+0.3}_{-0.3}$ | $2.00^{+0.02}_{-0.02}$ | 1.7/76 |
| 3 | $11.6^{+1.3}_{-1.5}$ | $0.26^{+0.02}_{-0.01}$ | $3.09^{+0.04}_{-0.04}$ | $10.6^{+0.3}_{-0.2}$ | $1.97^{+0.02}_{-0.02}$ | 0.6/75 |
|   | 0.62 | $0.32^{+0.08}_{-0.07}$ | $2.74^{+0.03}_{-0.03}$ | $10.5^{+0.2}_{-0.3}$ | $1.91^{+0.02}_{-0.02}$ | 1.0/76 |

[1] $N_H$ is the H I column density along the line of sight (in units of $10^{22}$ $cm^{-2}$; $T_b$ is the blackbody temperature; and $\alpha_1$ and $\alpha_2$ are the power-law photon indices below and above the break energy ($E_b$).

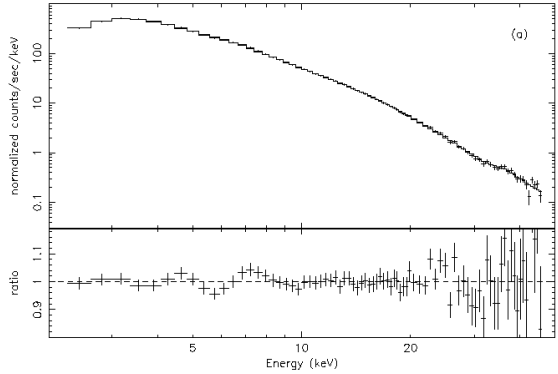

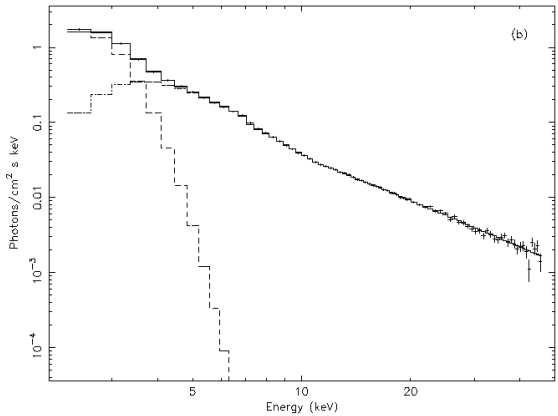

Figure 2: (a) Observed PCA spectrum of Cyg X-1 for the second observation. The solid-line histogram represents the best-fit model. (b) Unfolded X-ray spectrum for the same observation. Contributions from spectral components are shown individually.

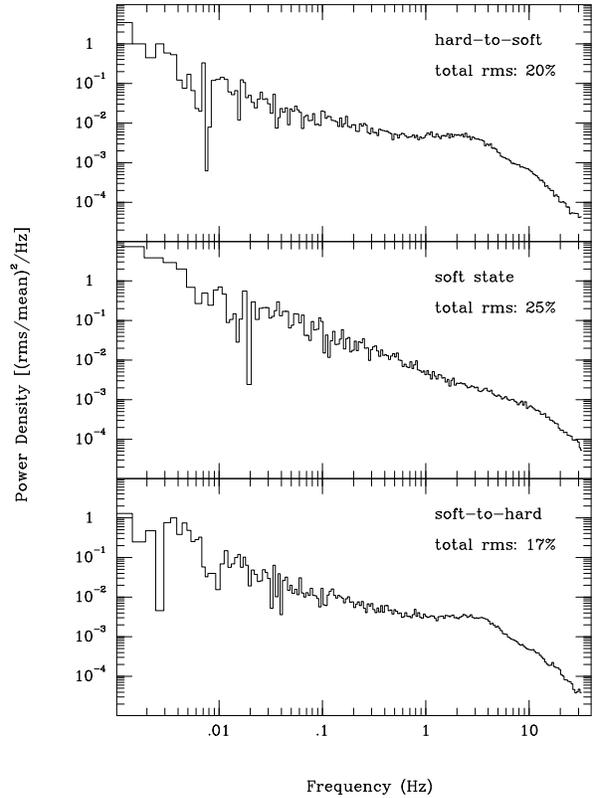

Figure 3: Evolution of the power density spectra derived from 22May96 (top), 16June96 (middle), and 12August96 (bottom) observations. Note that the integrated fractional rms variability shown is calculated in the frequency range of $\sim 1$ mHz-32 Hz.

3. TIMING ANALYSIS

From each of 3 observations we chose a contiguous stretch of data, and generated a PDS in the PCA energy band. The results were then logarithmically rebinned to reduce scatter at high frequencies. The resulting spectra are shown in Fig. 3. The PDSs have been corrected for instrumental artifacts due to electronic deadtime and very high energy events (Zhang et al. 1996d).

During the initial hard-to-soft transition (or the "settling period"; as represented by the first observation), the PDS can be characterized by a red noise component with a characteristic shape of 1/f at low frequencies, followed by a white noise component that extends to about 3 Hz, above which it is cut off. At higher frequencies, the PDS becomes power-law again, with a slope of roughly -2, i.e., "$1/f^2$". In the "true" soft state (as seen by the second observation), the 1/f noise completely dominates the PDS. Finally, during the soft-to-hard transition (typified by the last observation), as expected, the observed PDS becomes similar to that for the hard-to-soft transition: the white noise appears again, and is cut off at $\sim 3$ Hz. Accordingly, the 1/f noise is less dominant. No low-frequency 1/f noise has not been observed for Cyg X-1 in the normal hard state (van der Klis 1995). It is interesting to note that the total rms variability of the source appears to be higher in the soft state (see Fig. 3).

## 4. DISCUSSION

We interpret the soft blackbody component as the emission from a geometrically thin, optically thick cool accretion disk. The soft X-ray photons are Compton upscattered by a geometrically thick, optically thin corona surrounding the disk to produce the observed hard X-ray emission (Liang & Nolan 1984, and references therein). Then the spectrum can be approximated by a thermal component around the blackbody temperature, a power-law ($\alpha_1$) at energies just above, and a flatter power-law component ($\alpha_2$) at still higher energies before being cut off beyond $kT_e$, where $T_e$ is the electron temperature of the corona (Liang & Nolan 1984, and also see discussion by Ebisawa et al. 1996). A similar spectral shape was observed in the hard state by $ASCA$ (Ebisawa et al. 1996). In their case, the soft blackbody component had a lower temperature ($kT \simeq 0.1$ keV), and the broken power-law was flatter ($\alpha_1 = 1.92$ and $\alpha_2 = 1.71$), with a lower break energy ($\sim 3.4$ keV).

In the soft state the observed PDS shows a distinct 1/f component at low frequencies, which seems to be absent in the hard state. This component may be due to the superposition of random accretion "shots" with long lifetimes (see discussion by Belloni & Hasinger 1990). Theoretical models have been proposed to associate these shots with instabilities in the accretion disk (e.g., Mineshige, Takeuchi, & Nishimori 1994). Perhaps, the 1/f noise strengthens as the disk mass accretion rate increases. This would explain its dominance in the soft state (presumably with a higher disk accretion rate) and the absence in the low state. Our results show very nicely the evolution of the 1/f noise component during the spectral transitions (see Fig. 3).

The white noise (or flat) component has been seen in the hard state and during the spectral transitions. It may be due to statistical fluctuations in the mass accretion stream near the inner edge of the accretion disk where dynamical time scale is very short (compared to the frequency range that has been covered), similar to the "shot noise" in many electronic systems (van der Ziel 1986). The hot corona in the system can act as a low-pass filter that cuts off the white noise at some characteristic frequency to produce the observed "flat-top" PDS shape. In this model, the cutoff frequency is determined by the characteristic photon escape time. Then, the higher cutoff frequency observed in the high state ($\sim 3$ Hz, compared to $\sim 0.1$ Hz in the low state) can be explained by a smaller corona due to more efficient local cooling provided by a higher disk mass accretion rate. Because the corona is smaller in the soft state, the number of scatterings that an X-ray photon experiences is, on average, less, therefore the emerging hard X-ray spectrum is softer, which agrees qualitatively with the observations.

Based on both the spectral and timing results, Cui et al. (1997) proposed that the "true" soft state was not reached until the last observation, following a "settling period". This is now supported by our new results which show clearly that for the 16June1996 observation (the second in Table 2), the X-ray spectrum is soft and the PDS is dominated by the 1/f noise, very similar to that of the last observation in Cui et al. (1997), suggesting that the source had indeed settled down in the soft state (also see Cui, Focke, & Swank 1996). Here we also confirm that the low energy X-ray spectrum (i.e., $\alpha_1$) varies more significantly than the high energy spectrum ($\alpha_2$) during the transitions (Cui et al. 1997). In fact, only the low-energy spectrum is significantly softer in the soft state than that during the transitions. This could be evidence for a "stratified" corona (Skibo & Dermer 1995). The more stable high-energy component could be associated with the Comptonization in an inner and hotter region which is less sensitive to the soft photon flux produced in the outer disk, while the low-energy component may be due to Compton scattering in an outer, but slightly cooler region, which is more sensitive to the soft photon flux. It is the outer part of the corona that fluctuates between the two states. The integrated 2-60 keV fluxes were derived to be 3.3, 2.9, and $2.3 \times 10^{-8}\ erg\ cm^{-2}\ s^{-1}$, respectively, which is consistent with the fact that the bolometric X-ray luminosity changes little during the spectral transitions (Zhang et al. 1996c), contrary to the belief that the spectral states are determined by the total mass accretion rate.

We wish to thank every member of the $RXTE$ team for the success of the mission. This work is supported in part by NASA Contract NAS5-30612.